\documentclass{article}
\usepackage{graphicx} 

\title{Mixed-Precision Random Projection for RandNLA \\ on Tensor Cores}

\author{Hiroyuki Ootomo$^{1,\dag}$ and Rio Yokota$^{1,\dag\dag}$}

\date{$^1$ Tokyo Institute of Technology\\ $^\dag$ ootomo.h@rio.gsic.titech.ac.jp \\ $^{\dag\dag}$ rioyokota@gsic.titech.ac.jp}

\usepackage{amsmath,amssymb,amsfonts}
\usepackage{threeparttable}
\usepackage{multirow}
\usepackage{graphicx}
\usepackage{enumitem}
\usepackage{xcolor}
\usepackage{algorithm}
\usepackage{algpseudocode}
\usepackage{url}
\usepackage{colortbl}
\usepackage{multicol}
\usepackage[margin=3cm]{geometry}
\usepackage{hyperref}

\newtheorem{theorem}{Theorem}
\newtheorem{assumption}{Assumption}

\begin{document}

\maketitle

\begin{abstract}
Random projection can reduce the dimension of data while capturing its structure and is a fundamental tool for machine learning, signal processing, and information retrieval, which deal with a large amount of data today.
RandNLA (Randomized Numerical Linear Algebra) leverages random projection to reduce the computational complexity of low-rank decomposition of tensors and solve least-square problems.
While the computation of the random projection is a simple matrix multiplication, its asymptotic computational complexity is typically larger than other operations in a RandNLA algorithm.
Therefore, various studies propose methods for reducing its computational complexity.
We propose a fast mixed-precision random projection method on NVIDIA GPUs using Tensor Cores for single-precision tensors.
We exploit the fact that the random matrix requires less precision, and develop a highly optimized matrix multiplication between FP32 and FP16 matrices -- SHGEMM (Single and Half GEMM) -- on Tensor Cores, where the random matrix is stored in FP16.
Our method can compute Randomized SVD 1.28 times faster and Random projection high order SVD 1.75 times faster than baseline single-precision implementations while maintaining accuracy.
\end{abstract}


\section{Introduction}
Random projection is a robust tool for reducing data dimension and compressing data while preserving its structure by projecting it onto a lower-dimensional subspace.
It is used for various methods such as machine learning \cite{chiu_self-supervised_2022,vinh_training_2016,fern_random_2003,chen_efficient_2018,fradkin_experiments_2003}, computer vision \cite{kim_bilinear_2015,tsagkatakis_random_2009,mccane_random_2014}, database search \cite{achlioptas_database-friendly_2003,hegde_random_2007}, and other applications \cite{nabil_random_2017,zhang_projection-based_nodate}.
As the amount of data we process in such compressed formats increases, the performance of random projection becomes critical.
RandNLA (Randomized Numerical Linear Algebra) \cite{drineas_randnla_2016} is a group of numerical algorithms that leverage the random projection and other randomized methods for reducing the computational complexity of low-rank approximation of tensors \cite{halko_finding_2011,shabat_randomized_2018,erichson_randomized_2019,minster_randomized_2020,ahmadi-asl_randomized_2021,che_randomized_2021}, and least-squares methods \cite{rokhlin_fast_2008}.
For instance, Randomized SVD (Singular Value Decomposition) is a fast low-rank approximation algorithm for matrices with predetermined approximation rank \cite{halko_finding_2011}.
While the low-rank approximation of a matrix using SVD is a fundamental operation, the computational complexity of SVD is large.
The Randomized SVD and its variants reduce the complexity and are used for image and data compression \cite{erichson_compressed_2017}, matrix completion \cite{feng_faster_2018}, digital watermarking \cite{anand_cloud_2021,singh_secdh_2022}, and other research fields \cite{yang_low_2018,zhang_randomized_2012,li_large-scale_2015,wang_large-scale_2022,intawichai_missing_2022}.
They can be used through {\tt scikit-learn} Python packages \cite{pedregosa_scikit-learn_2011}, for instance.
Although the random projection is a simple multiplication of an input matrix and a random matrix, its asymptotic computational complexity is larger than other parts, such as the SVD and QR factorization of the small and skinny matrices, in RandNLA algorithms.
Therefore, various studies focus on reducing the computational complexity of the random projection using a structured random matrix \cite{halko_finding_2011}, sparse random matrix \cite{achlioptas_database-friendly_2003,li_very_2006}, and Hadamard matrix \cite{menon_fast_2016}.
However, to the best of our knowledge, there are very few studies on using the latest hardware features to accelerate the random projection.

In response to the high demand for fast matrix multiplication computations in deep learning, several specialized computing units for matrix multiplication have been developed, \textit{e.g.} NVIDIA Tensor Core \cite{corporation_nvidia_2020}, AMD Matrix Core \cite{amd_amd_2021}, Google TPU \cite{jouppi_-datacenter_2017}, and Preferred Networks MN-Core \cite{makino_present_2021}.
NVIDIA Tensor Core is a mixed-precision matrix multiplication and addition computing unit, and its theoretical peak performance is 312 TFlop/s on A100 GPU.
Although the input data type of two matrices for multiplication is stored in low-precision (FP16 or TF32), the computation inside is performed in FP32, and we can obtain the resulting matrix in FP32.
However, the resulting accuracy degrades when computing a single-precision matrix multiplication on Tensor Cores since we need to convert input FP32 matrices to low-precision.
To recover the accuracy, Markidis \textit{et al.} propose a method for recovering the accuracy using a compensated summation.
Their method splits each input FP32 matrix into a sum of two FP16 matrices and sums up the result of the multiplication of each sub-divided matrix on Tensor Cores \cite{markidis_nvidia_2018}.
Although their method can recover the accuracy theoretically, it is still worse than single-precision in practice \cite{ootomo_recovering_2022}.
Our previous work shows that the rounding mode used inside Tensor Cores (RZ) is the cause of the accuracy degradation.
We propose a method, TCEC-SGEMM (Tensor Core with Error Correction), that avoids the accumulation inside Tensor Cores by performing the accumulation on the FP32 SIMT Cores.
We have demonstrated that our method outperforms the theoretical peak performance of FP32 SIMT Cores on A100 GPU while the accuracy is at the same level.
As another splitting approach for computing high-precision GEMM on Tensor Cores, Mukunoki \textit{et al.} use the Ozaki scheme \cite{ozaki_error-free_2012} and show that their method can compute double-precision matrix multiplication on Tensor Cores \cite{mukunoki_dgemm_2020}.
The main difference between the Ozaki scheme and TCEC-SGEMM is that while the former method splits the mantissa of input matrices by thresholds shared by rows or columns and conducts error-free matrix multiplications of sub-divided matrices, the latter method splits the mantissa by element local thresholds and all matrix multiplications of sub-divided matrices introduces rounding error.
Although Mukunoki \textit{et al.} demonstrate that their method can compute double-precision equivalent matrix multiplication faster than FP64 theoretical peak performance on NVIDIA TITAN RTX GPU, single-precision multiplication is slower.

We propose a mixed-precision random projection method on Tensor Cores to improve the random projection throughput for a single-precision tensor.
In previous research, the data type of the Gaussian random matrix in the random projection is the same as the input matrix (FP32) since it can leverage highly optimized GEMM implementations for each processor.
On the other hand, we use an FP16 random matrix for the random projection since our GEMM method -- {\bf SHGEMM} (Single and Half GEMM) -- can multiply it to an FP32 input matrix faster than SGEMM while maintaining accuracy.
Therefore, using a low-precision random matrix allows us to reduce the memory usage for the random matrix and improve the computational throughput.
We show that the method improves the throughput of RandNLA algorithms while maintaining accuracy as well.

The main contributions of this paper are highlighted below.
\begin{itemize}
    \item We investigate the properties of the low-precision representation of the Gaussian random matrix and show that we can use it for the random projection.
    \item We implement SHGEMM, a matrix multiplication between FP32 and FP16 matrices using Tensor Cores.
    We show its rounding error analysis, accuracy, and throughput and investigate the throughput bottleneck.
    \item We evaluate the low-precision Gaussian random projection and SHGEMM in two RandNLA algorithms.
    We improve the throughput of Randomized SVD by $1.28$ times and Random projection HOSVD by $1.49$ times compared to baseline FP32 implementations while maintaining accuracy.
\end{itemize}

\section{Background}
\subsection{Low-rank approximation using SVD}
For a complex matrix $\mathbf{A} \in \mathbb{C}^{m, n}$, SVD (Singular Value Decomposition) decompose $\mathbf{A}$ as a multiplication of three matrices as $\mathbf{A}=\mathbf{U}\cdot\mathbf{\Sigma}\cdot\mathbf{V}^\top$.
The matrices $\mathbf{U}$ and $\mathbf{V}$ are unitary matrices, and $\mathbf{\Sigma}$ is a diagonal matrix where diagonal elements $\sigma_1, \cdots, \sigma_k$ are singular values of $\mathbf{A}$ and $k$ is the rank of $\mathbf{A}$.
The $p$-rank approximation of $\mathbf{A}$ by SVD (truncated SVD; tSVD) can be calculated by the truncation of $[p+1:k]$ column vectors of $\mathbf{U}, \mathbf{V}$ and the diagonal elements of $\mathbf{\Sigma}$ as follows:
\begin{align*}
    \mathbf{A} &\sim \hat{\mathbf{A}} = \hat{\mathbf{U}} \cdot \mathbf{\Sigma}_1 \cdot \hat{\mathbf{V}^\top}
\end{align*}
where
\begin{align*}
    \hat{\mathbf{U}} \leftarrow \mathbf{U}_{[1:p]},
    \hat{\mathbf{V}} \leftarrow \mathbf{V}_{[1:p]},
    \mathbf{\Sigma}_1 \leftarrow \text{diag}(\sigma_1, \cdots, \sigma_p).
\end{align*}
The Eckart bounds the approximation accuracy–Young theorem \cite{eckart_approximation_1936}.
\begin{theorem}[Eckart–Young theorem]
\label{the:eym}
\begin{equation}
    ||\mathbf{A} - \hat{\mathbf{A}}||_F = ||\mathbf{\Sigma}_2||_F,
\end{equation}

where $\mathbf{\Sigma}_2 = \text{diag}(\sigma_{p+1},\cdots,\sigma_k)$ and $||\cdot||_F$ denotes the Frobenius norm.

\end{theorem}
Since the computational complexity of SVD for an $m \times n$ matrix is $\mathcal{O}(mn\cdot \text{min}(m, n))$ and large, we do not compute the full SVD of the input matrix when the approximation rank is already known.
Instead, we use an algorithm based on the rank-revealing QR decomposition \cite{gu_efficient_2012} to compute the same approximation in $\mathcal{O}(mnp)$.

When the singular values decay slowly, we can use a power scheme that changes the input matrix to $\hat{\mathbf{A}}$ calculated as follows:
\begin{equation*}
    \hat{\mathbf{A}} = \left(\mathbf{AA}^\top\right)^q \mathbf{A} = \mathbf{U}\cdot\mathbf{\Sigma}^{2q+1}\cdot\mathbf{V}^\top,
\end{equation*}
where $q$ is the number of the power iterations.

\subsection{Random Projection in RandNLA}
In this section, we take Randomized SVD as an example to explain the random projection method in RandNLA.
The Randomized SVD algorithm can be used when the approximation rank is given.
In a $p$-rank approximation by Randomized SVD, we calculate a matrix $\mathbf{Y} \in \mathbb{R}^{m \times p}$ as follows:
\begin{equation}
    \label{eq:rp-vec}
    \mathbf{Y}=\mathbf{A\Omega}
\end{equation}
where $\mathbf{\Omega} \in \mathbb{R}^{n \times p}$ is a random matrix.
Since the column vectors of $\mathbf{Y}$ are the linear combinations of the column vectors of $\mathbf{A}$, these two matrices share the orthonormal vectors.
Therefore, an orthogonal matrix $\mathbf{Q}$ obtained by a QR factorization of $\mathbf{Y}$, for instance, is also the orthonormal vectors of $\mathbf{A}$.
Thus, $\mathbf{A}$ is approximated as follows:
\begin{equation}
    \label{eq:qqta}
    \mathbf{A} \sim \mathbf{Q}\mathbf{Q}^\top\mathbf{A}.
\end{equation}
Then, we can calculate the $p$-rank approximation by computing the SVD of $\mathbf{Q}^\top \mathbf{A}$ as follows:
\begin{align*}
    \hat{\mathbf{U}}', \mathbf{\Sigma}_1, \hat{\mathbf{V}} \leftarrow \text{SVD}\left(\mathbf{Q}^\top\mathbf{A}\right),
    \hat{\mathbf{U}} \leftarrow \mathbf{Q}\hat{\mathbf{U}}'.
\end{align*}
The computation of Eq. (\ref{eq:rp-vec}) is called {\bf random projection}.
In practice, we set $p$ as $\hat{p} = p + s$ where $s (\geq 2)$ is an oversampling parameter.

When using a Gaussian random matrix $\mathbf{G}$ where each element is generated with $\mathcal{N}(0, 1)$, the accuracy of the random projection is bounded as follows \cite{halko_finding_2011}:
\begin{equation}
    \label{eq:qqta-acc}
    \mathbb{E}_\mathbf{G}||\mathbf{A} - \mathbf{QQ}^\top\mathbf{A}||_F \leq \left(\sqrt{1 + p / (s-1)}\right) ||\mathbf{\Sigma}_2||_F,
\end{equation}
where $\mathbb{E}_{X}$ is the expectation with respect to $X$, and $||\cdot||_F$ is the Frobenius norm.

\begin{algorithm}[t]
\caption{Randomized SVD}\label{alg:rsvd}
\begin{algorithmic}[1]
\Require Input matrix $\mathbf{A}\in\mathbb{C}^{m \times n}$, $\mathbf{\Omega}\in \mathbb{R}^{n \times \hat{p}}$, Target rank $p \in\mathbb{N}$
\Ensure $\hat{\mathbf{U}}, \mathbf{\Sigma}_1, \hat{\mathbf{V}}, s.t. \hat{\mathbf{A}}=\hat{\mathbf{U}}\cdot\mathbf{\Sigma}_1\cdot\hat{\mathbf{V}}^{\top}$
\State $\mathbf{Y} \gets \mathbf{A}\cdot\mathbf{\Omega}$
\State $\mathbf{Q}, \mathbf{R} \gets \text{QR}(\mathbf{Y})$
\State $\mathbf{B} \gets \mathbf{Q}^\top \cdot \mathbf{A}$
\State $\hat{\mathbf{U}}', \mathbf{\Sigma}_1, \hat{\mathbf{V}} \gets \text{tSVD}(\mathbf{B}, rank=p)$
\State $\hat{\mathbf{U}} \gets \mathbf{Q}\cdot\hat{\mathbf{U}}'$
\end{algorithmic}
\end{algorithm}

We show the algorithm of the Randomized SVD in Algorithm \ref{alg:rsvd}.
In the algorithm, the asymptotic computational complexity of the QR factorization (Line 2) and SVD (Line 4) are $\mathcal{O}(mp^2)$ and $\mathcal{O}(np^2)$, respectively, and the random projection (Line 1) is $\mathcal{O}(mnp)$.
Therefore, since $p \ll m, n$ typically, the random projection is the most expensive computation in the algorithm, and the total asymptotic complexity of the Randomized SVD is $\mathcal{O}(mnp)$.
Although this is the same as the complexity of a tSVD algorithm based on the rank-revealing QR decomposition, the Randomized SVD has been shown to be experimentally faster than that algorithm \cite{halko_finding_2011}.
Furthermore, to reduce the complexity of the random projection, there are some studies on using random matrices instead of the Gaussian random matrix.
For instance, when using \textit{subsampled random Fourier transform}, the computational complexity is $\mathcal{O}(mn \cdot \text{log}(p))$ \cite{halko_finding_2011}.
We can also use sparse random matrices for the random projection, although they were not originally proposed in the context of RandNLA.
These matrices are sparse from the perspective of non-zero elements and the choice of element values.
For instance, Achlioptas proposes a random matrix $\mathbf{\Omega} \in \mathbb{R}^{n \times q}$ where the $(i, j)$ element is decided as follows \cite{achlioptas_database-friendly_2003}:
\begin{equation}
    \label{eq:sp-rand}
    \mathbf{\Omega}_{(i, j)} = \sqrt{s} \times \left\{
\begin{array}{ll}
+1 & \text{with probability 1/2s} \\
0  & \text{with probability 1 - 1/s} \\
-1 & \text{with probability 1/2s}
\end{array}
\right.
\end{equation}
for $s=1$ and $3$.
Li \textit{et al.} propose a {\it very sparse random projection} where $s \gg 3$, especially $s = \sqrt{n}$ or $n/\log n$ \cite{li_very_2006}.

\subsection{Single-precision GEMM emulation on Tensor Cores}
\begin{figure}
    \centering
    \includegraphics[width=0.6\linewidth]{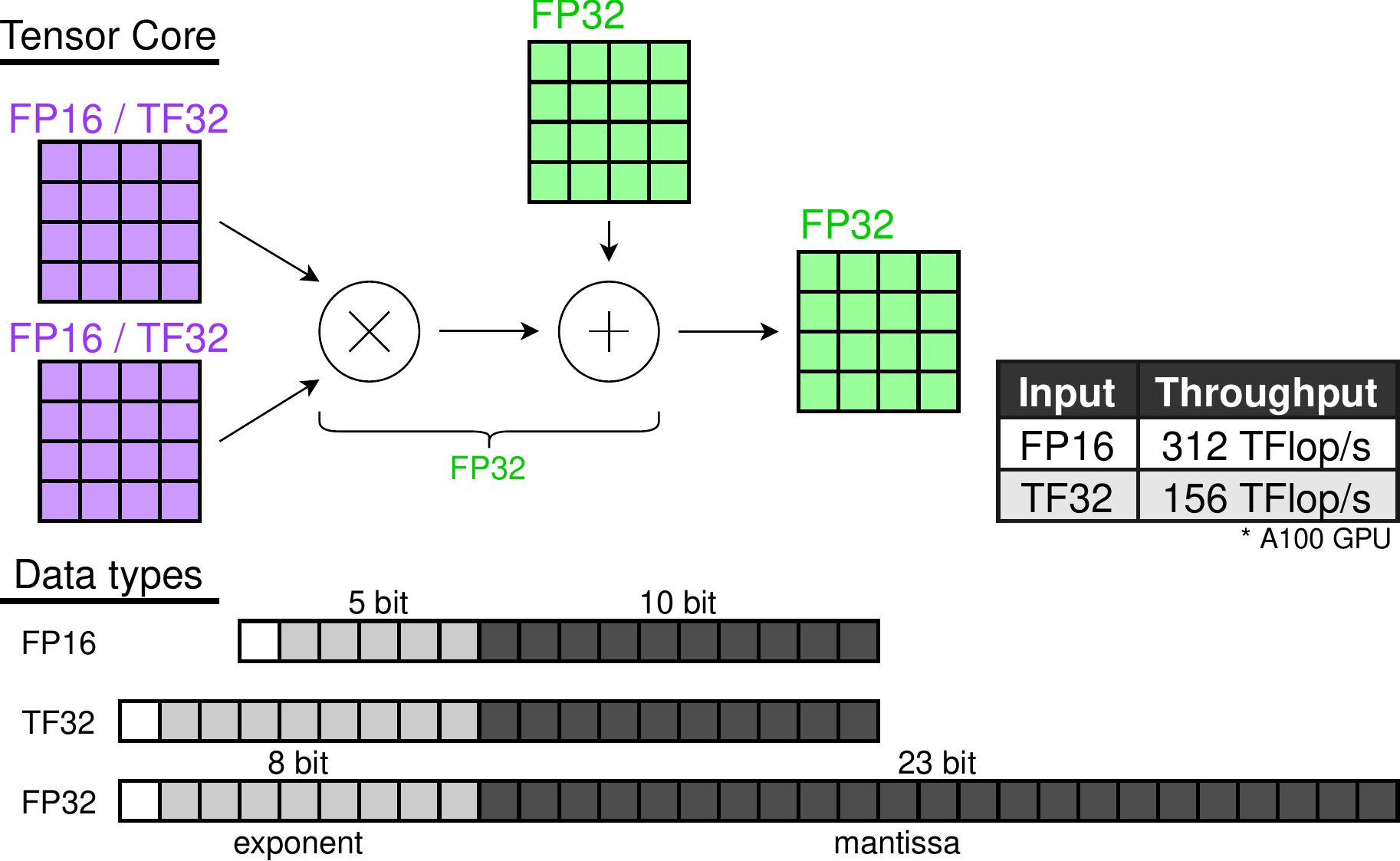}
    \caption{The input, output, and computation precision of Tensor Cores we use.}
    \label{fig:tensorcore}
\end{figure}
NVIDIA Tensor Cores are mixed-precision computing units for fixed-size matrix multiplications and additions on NVIDIA GPUs.
When computing a large matrix multiplication on Tensor Cores, we split the input matrices and sum up the resulting matrices.
The data type of input matrices to Tensor Cores is limited to low-precision, FP16, TF32, etc.
Although the multiplication and addition are performed in high-precision, such as FP32, as shown in Fig. \ref{fig:tensorcore}, the accuracy degrades when computing a single-precision GEMM on Tensor Cores since the input matrices have to be converted to low-precision.
Our previous work (TCEC-SGEMM method) recovers this accuracy through compensated summation and the avoidance of rounding inside Tensor Cores, RZ (Round to Zero) \cite{ootomo_recovering_2022}.
We have demonstrated that the method can outperform the FP32 SIMT Core peak performance while the accuracy is at the same level as single-precision.
In our method, the single-precision matrix multiplication $\mathbf{C}_\text{F32} \leftarrow \mathbf{A}_\text{F32}\cdot\mathbf{B}_\text{F32}$ is approximately computed as follows:
\begin{align}
    \mathbf{A}_\text{low} &\leftarrow \text{toLow}\left(\mathbf{A}_\text{F32}\right) \\
    \Delta \mathbf{A}_\text{low} &\leftarrow \text{toLow}\left(\left(\mathbf{A}_\text{F32} - \text{toF32}\left(\mathbf{A}_\text{low}\right)\right)\times 2^{11}\right) \\
    \mathbf{B}_\text{low} &\leftarrow \text{toLow}\left(\mathbf{B}_\text{F32}\right) \\
    \Delta \mathbf{B}_\text{low} &\leftarrow \text{toLow}\left(\left(\mathbf{B}_\text{F32} - \text{toF32}\left(\mathbf{B}_\text{low}\right)\right)\times 2^{11}\right) \\
    \label{eq:tcec-5}
    \mathbf{C}_\text{F32} &\sim \mathbf{A}_\text{low}\cdot\mathbf{B}_\text{low} + (\Delta\mathbf{A}_\text{low}\cdot\mathbf{B}_\text{low} + \mathbf{A}_\text{low}\cdot\Delta\mathbf{B}_\text{low})\times2^{-11},
\end{align}
where toLow converts an FP32 matrix to low-precision, FP16 or TF32, and toF32 converts a low-precision matrix to FP32.
Since the matrix multiplications in Eq. (\ref{eq:tcec-5}) are computed on Tensor Cores in high precision, this method manages to compute the matrix multiplication in single precision.
And we compute parts of additions inside $\mathbf{A}_\text{low}\cdot\mathbf{B}_\text{low}$ in Eq. (\ref{eq:tcec-5}) on FP32 SIMT Cores to use Round to Nearest (RN) for the rounding.
We denote a Tensor Core which takes FP16 matrices as ``FP16 Tensor Core" and one which takes TF32 matrices as ``TF32 Tensor Core".

\section{Random projection by low-precision Gaussian random matrix}
\subsection{Floating point value representation of a Gaussian random value}
The theoretical error bound of the random projection by Halko \textit{et al.}, Eq. (\ref{eq:qqta-acc}), is proved mathematically under the implicit assumption that the Gaussian random value $g \in \mathcal{N}(0, 1)$ has infinite precision.
On the other hand, we represent $g$ as a floating point value format eXmY, which has X bit of exponent and Y bit of mantissa, and denote it as $g_\text{eXmY}$.
For instance, FP32 is e8m23 and FP16 is e5m10.
Furthermore, we use RN for the rounding, which is used by default in many modern processors, including NVIDIA GPUs.
Due to the rounding error, the relation between $g$ and $g_\text{eXmY}$ is as follows.
\begin{equation}
    \label{eq:g-fp-error}
    g_\text{eXmY} = g + \Delta g
\end{equation}
where $\Delta g$ is the rounding error.
In the case of RN, the following inequality is satisfied.
\begin{equation}
    \label{eq:RN-range}
    - \text{exp\_of}(g_\text{eXmY}) \times u_Y \leq \Delta g \leq \text{exp\_of}(g_\text{eXmY}) \times u_Y,
\end{equation}
where $\text{exp\_of}(\cdot)$ is the exponent of the floating point value and $u_Y = 2^{-(Y+1)}$.

\subsection{Properties of low-precision Gaussian random value}
The property of Gaussian random values represented in floating point varies among its formats.
We examine the following three properties.
\begin{enumerate}
    \item The overflow and underflow probability.
    \item The number of elements within a specific range of the Gaussian distribution.
    \item The mean and variance.
\end{enumerate}

\begin{table}[]
\centering
\begin{tabular}{lccc}
format        & $p^\text{of}_{\tt eXmY}$                                                                            & $p^\text{not-normalized}_{\tt eXmY}$                              & $p^\text{uf}_{\tt eXmY}$                                  \\
\hline
\cellcolor[gray]{.9} FP8\_1 (e4m3)    & \multirow{6}{*}{\begin{tabular}[c]{@{}c@{}}$\ll 2-2\Phi(2^3)$\\ $< 1 \times 10^{-12}$\end{tabular}} &\cellcolor[gray]{.9}  $6\times10^{-3}$                                                  &\cellcolor[gray]{.9}  $8\times10^{-4}$                                             \\
FP8\_2 (e5m2)    &                                                                                                       & $2\times10^{-5}$                                                  & $6\times10^{-6}$                                             \\
\cellcolor[gray]{.9} FP16 (e5m10)  &                                                                                                       &\cellcolor[gray]{.9}  $2\times10^{-5}$                                                  &\cellcolor[gray]{.9}  $2\times10^{-8}$                                             \\
bfloat (e8m7) &                                                                                                       & \multicolumn{2}{c}{\multirow{3}{*}{\begin{tabular}[c]{@{}c@{}}$\ll 2 \Phi(2^{-45}) - 1$\\ $< 2\times 10^{-12}$\end{tabular}}} \\
\cellcolor[gray]{.9} TF32 (e8m10)  &                                                                                                       & \multicolumn{2}{c}{}                                                                                                             \\
FP32 (e8m23)  &                                                                                                       & \multicolumn{2}{c}{}                                                                                                             \\
\hline
              & $N^{1\sigma}_\text{\tt eXmY}$                                                                      & $N^{2\sigma}_\text{\tt eXmY}$                                  & $N^{4\sigma}_\text{\tt eXmY}$                             \\
\hline
\cellcolor[gray]{.9}FP8\_1 (e4m3)    & \cellcolor[gray]{.9}111                                                                                                   & \cellcolor[gray]{.9}127                                                               & \cellcolor[gray]{.9}143                                                          \\
FP8\_2 (e5m2)    & 119                                                                                                   & 127                                                               & 135                                                          \\
\cellcolor[gray]{.9}FP16 (e5m10)  & \cellcolor[gray]{.9}30,719                                                                                                & \cellcolor[gray]{.9}32,767                                                            & \cellcolor[gray]{.9}34,815                                                       \\
bfloat (e8m7) & 32,511                                                                                                & 32,767                                                            & 33,023                                                       \\
\cellcolor[gray]{.9}TF32 (e8m10)  & \cellcolor[gray]{.9}260,095                                                                                               & \cellcolor[gray]{.9}262,143                                                           & \cellcolor[gray]{.9}264,191                                                      \\
FP32 (e8m23)  & 2,130,706,431                                                                                         & 2,147,483,647                                                     & 2,164,260,863                                               
\end{tabular}
\caption{The properties of Gaussian random values represented in floating point values eXmY. {\bf Top:} $p_\text{eXmY}^\text{of}, p_\text{eXmY}^\text{uf}$ and $p_\text{eXmY}^\text{not-normalized}$ are the probabilities of overflow, underflow, and being denormalized values or underflowed, respectively. {\bf Bottom:} The number of floating point values within $\sigma$, $2\sigma$, and $4\sigma$ ranges.}
\label{tab:eXmY-gaussian}
\end{table}
\subsubsection{The overflow and underflow probability}
If a Gaussian random value overflows with non-negligible probability, we can not compute the random projection correctly.
On the other hand, it can be computed even if the underflow occurs with high probability since this is just a sparse random matrix \cite{li_very_2006}.
We investigate these probabilities theoretically.

The maximum positive value $\text{max}_\text{eXmY}$ that can be represented as eXmY is calculated as follows:
\begin{align}
    \text{max}_\text{eXmY} &= 0b0'{\underbrace{11\cdots 10}_\text{X bit}} ' \underbrace{11\cdots 1}_\text{Y bit} \\
    &=2^{2^{X}-2-\text{bias}(X)} \times \left(1 - 2^{-(Y+1)}\right) \\
    &=2^{2^{X-1}-1} \times \left(1 - 2^{-(Y+1)}\right),
\end{align}
where $\text{bias}(X)=2^{X-1}-1$ is the exponent bias of eXmY.
Therefore, the overflow probability $p_\text{eXmY}^\text{of}$, in other words, the probability that a value larger than $\text{max}_\text{eXmY}$ appears is calculated as follows:
\begin{equation}
    p_\text{eXmY}^\text{of} = 2\times \left(1 - \Phi(\text{max}_\text{eXmY})\right)
\end{equation}
where $\Phi(\cdot)$ is the probability density function of the Gaussian distribution $\mathcal{N}(0, 1)$.
On the other hand, the minimum positive values $\text{min}_\text{eXmY}^\text{normalized}$ and $\text{min}_\text{eXmY}^\text{denormalized}$ that can be represented as a normalized and denormalized value of eXmY respectively are calculated as follows:
\begin{align*}
    \text{min}_\text{eXmY}^\text{normalized} &= 0b0'{\underbrace{00\cdots 01}_\text{X bit}} ' \underbrace{00 \cdots 0}_\text{Y bit} \\
    &= 2^{1 - \text{bias}(X)}
    =2^{2-2^{X-1}}
\end{align*}
\begin{align*}
    \text{min}_\text{eXmY}^\text{denormalized} &= 0b0'{\underbrace{00\cdots 00}_\text{X bit}} ' \underbrace{00 \cdots 01}_\text{Y bit} \\
    &= 2^{- \text{bias}(X)} \times 2^{-Y}
    = 2^{1-2^{X-1}} \times 2^{-Y}.
\end{align*}
Thus, the underflow probability $p_\text{eXmY}^\text{uf}$ and denormalized probability $p_\text{eXmY}^\text{not-normalized}$ are calculated as follows:
\begin{align*}
    p_\text{eXmY}^\text{uf} &= 2\times(\Phi(\text{min}_\text{eXmY}^\text{denormalized}) - 1/2) \\
    p_\text{eXmY}^\text{not-normalized} &= 2\times(\Phi(\text{min}_\text{eXmY}^\text{normalized}) - 1/2).
\end{align*}
We show concrete values of $p_\text{eXmY}^\text{of}, p_\text{eXmY}^\text{uf}$ and $p_\text{eXmY}^\text{not-normalized}$ for some floating point formats in the top of Table \ref{tab:eXmY-gaussian}.
Since the number of elements in a random matrix for random projection is typical up to $\sim 10^8$ ($n < 10^5, p < 10^3$), the overflow probability is negligible when the exponent length X is larger than 3.

\subsubsection{The number of elements within $2^s \sigma$ range}
If the number of elements within a specific range of the Gaussian distribution is too small, the Gaussian random matrix might not have full rank, albeit with a very low probability, which is assumed in the proof of Eq. (\ref{eq:qqta-acc}).
As an extreme example, if only one value can appear, the rank of the random matrix is 0 or 1.
We investigate the number of elements within the $2^s \sigma$ range of the Gaussian distribution.

A positive normalized number element $v_\text{eXmY}$ within $2^s \sigma$ range of the Gaussian distribution $\mathcal{N}(0, \sigma=1)$ satisfies the following inequality:
\begin{equation}
    \text{min}_\text{eXmY}^\text{normalized} \leq v_\text{eXmY} < 0b0'{\underbrace{**\cdots **}_{=s + \text{bias}(X)}} ' \underbrace{00 \cdots 0}_\text{Y bit}.
\end{equation}
Therefore, the number of normalized number elements $N_\text{eXmY}^{2^s}$ within $2^s \sigma$ including 0 is calculated as follows:
\begin{align}
    N_\text{eXmY}^{2^s\sigma} &= 2 \times (s + \text{bias}(X) + 1)\times 2^Y + 1.
\end{align}
We show the concrete values of $N_\text{eXmY}^{2^s}$ for some floating point formats at the bottom of Table \ref{tab:eXmY-gaussian}.
Although the number of elements in low-precision formats, such as FP16, is fewer than one in FP32, it is still larger than the existing sparse random projection methods, typically 3.

\subsubsection{The mean and variance}
When using RN for rounding, the mean of $g_\text{eXmY}$ is $0$ since the floating point value is symmetric with respect to the sign.
On the other hand, the variance $\alpha_Y = \mathbb{E}_g({g_\text{eXmY}}^2) - (\mathbb{E}_g(g_\text{eXmY}))^2=\mathbb{E}_g({g_\text{eXmY}}^2)$, where $\mathbb{E}_x (y)$ is the mean of $y$ for $x$, is not $1$ even if the value generated with $\mathcal{N}(0, 1)$ and rounded.
\begin{figure}
    \centering
    \includegraphics[width=0.6\linewidth]{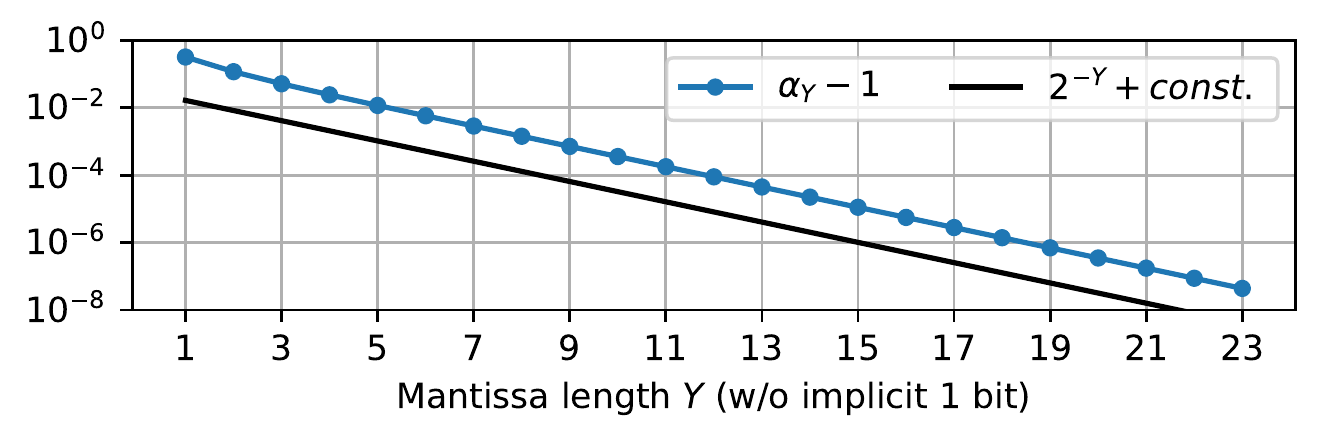}
    \caption{The variance of the Gaussian distribution represented in eXmY.}
    \label{fig:var-fp}
\end{figure}
The variance $\alpha_Y$ is calculated as follows:
\begin{align*}
    \alpha_Y &= \mathbb{E}_g({g_\text{eXmY}}^2) \\
    &= \sum_{v \in \text{eXmY}} v^2 \left(\Phi(v + \text{exp\_of}(v)u_Y) - \Phi(v - \text{exp\_of}(v))u_Y\right)
\end{align*}
While this equation can not be calculated analytically, we perform a numerical computation, as shown in Fig. \ref{fig:var-fp}.
The variance exponentially approaches $1$ for the mantissa length $Y$.

While Eq. (\ref{eq:qqta-acc}) is the error bound when using $\mathcal{N}(0, 1)$, we show the bound can be applied for any variance $\alpha$ as long as it is not too small.
The proof of Eq. (\ref{eq:qqta-acc}) comes as a proof of the following equation:
\begin{equation}
    \label{eq:sgg}
    \mathbb{E}||\mathbf{\Sigma}_2 \mathbf{G}_2 \mathbf{G}_1^\dag||_F^2 = k/(p-1) \cdot ||\mathbf{\Sigma}_2||^2_F,
\end{equation}
where $\Sigma_2$ is $\text{diag}(\sigma_{p+1}, \cdots, \sigma_k)$, $\mathbf{G}_1, \mathbf{G}_2$ are $k \times k$ and $k \times (k-p)$ Gaussian random matrices respectively, and $\cdot^\dag$ is a pseudo inverse of a matrix.
This equation can be proven by following two theorems in \cite{halko_finding_2011} under the assumption that a Gaussian random matrix has full rank.

\begin{theorem}
\label{thm:sgt}
For any matrix $\mathbf{S} \in \mathbb{C}^{n_1 \times n_2}, \mathbf{T} \in \mathbb{C}^{n_3 \times n_4}$ and a $\mathcal{N}(0, 1)$-Gaussian matrix $\mathbf{G} \in \mathbb{C}^{n_2 \times n_3}$,
\begin{equation}
    \mathbb{E}||\mathbf{S}\mathbf{G}\mathbf{T}||_F^2 = ||\mathbf{S}||_F^2 ||\mathbf{T}||_F^2.
\end{equation}
\end{theorem}
\begin{theorem}
\label{thm:g-inv}
For a $\mathcal{N}(0, 1)$-Gaussian random matrix $\mathbf{G} \in \mathbb{C}^{k \times (k + p)}$,
\begin{equation}
    \mathbb{E}||\mathbf{G}^\dag||_F^2 = k / (p-1).
\end{equation}
\end{theorem}
From these theorems, Eq. (\ref{eq:sgg}) is proved as follows.
\begin{align}
    \label{eq:sgg-0}
    \mathbb{E}_{\mathbf{G}_1, \mathbf{G}_2}||\mathbf{\Sigma}_2 \mathbf{G}_2 \mathbf{G}_1^\dag||_F^2 &= \mathbb{E}_{\mathbf{G}_1}\left(\mathbb{E}_{\mathbf{G}_2}\big[||\mathbf{\Sigma}_2 \mathbf{G}_2 \mathbf{G}_1^\dag||_F^2 \big\rvert \mathbf{\mathbf{G}_1}\big] \right) \\
    \label{eq:sgg-1}
    &= \mathbb{E}_{\mathbf{G}_1}\left(||\mathbf{\Sigma}_2||_F^2 ||\mathbf{G}_1^\dag||_F^2\right)  & \because \text{Theorem \ref{thm:sgt}}\\
    \label{eq:sgg-2}
    &= ||\mathbf{\Sigma}_2||_F^2 \cdot \mathbb{E}_{\mathbf{G}_1}||\mathbf{G}_1^\dag||_F^2 \\
    \label{eq:sgg-3}
    &= k/(p-1) \cdot ||\mathbf{\Sigma}_2||_F^2, & \because \text{Theorem \ref{thm:g-inv}}
\end{align}
where $\mathbb{E}[X|Y]$ is a conditional expectation of $X$ given $Y$.
\begin{flushright}
$\square$
\end{flushright}

To show that Eq. (\ref{eq:qqta-acc}) is also true when using $\mathcal{N}(0, \alpha)$, it suffices to show Eq. (\ref{eq:sgg}) is true for any variance $\alpha$.
Theorem \ref{thm:sgt} and Theorem \ref{thm:g-inv} are extended as follows.

\begin{theorem}[Extension of Theorem \ref{thm:sgt}]
\label{thm:sgt-fp}
For any matrix $\mathbf{S} \in \mathbb{C}^{n_1 \times n_2}, \mathbf{T} \in \mathbb{C}^{n_3 \times n_4}$ and a $\mathcal{N}(0, \alpha)$-Gaussian matrix $\mathbf{G}_\alpha \in \mathbb{C}^{n_2 \times n_3}$,
\begin{equation}
    \mathbb{E}||\mathbf{S}\mathbf{G}_\alpha\mathbf{T}||_F^2 = \alpha ||\mathbf{S}||_F^2 ||\mathbf{T}||_F^2.
\end{equation}
\end{theorem}

\begin{theorem}[Extension of Theorem \ref{thm:g-inv}]
\label{thm:g-inv-fp}
For a $\mathcal{N}(0, \alpha)$-Gaussian random matrix $\mathbf{G}_\alpha \in \mathbb{C}^{k \times (k + p)}$,
\begin{equation}
    \mathbb{E}||{\mathbf{G}_\alpha}^\dag||_F^2 = k / (p-1) \cdot \alpha^{-1}.
\end{equation}
\end{theorem}

We apply these theorems in Eq. (\ref{eq:sgg-0}-\ref{eq:sgg-3}) and obtain the same equation as Eq. (\ref{eq:sgg}).
\begin{flushright}
$\square$
\end{flushright}

\textit{Proof of Theorem \ref{thm:sgt-fp}.}
For $\mathbf{M}:=\mathbf{SGT}$ and $\mathbf{M}_\alpha:=\mathbf{S}\mathbf{G}_\alpha \mathbf{T}$, it suffices to show $\mathbb{E}|{\mathbf{M}_\alpha}_{(i, j)}|^2 = \alpha \mathbb{E}|\mathbf{M}_{(i, j)}|^2$, where $\cdot_{(i, j)}$ is the $(i, j)$ element of the matrix.
The elements $\mathbf{M}_{(i, j)}$ and ${\mathbf{M}_\alpha}_{(i, j)}$ are calculated as follows:
\begin{align*}
    \mathbf{M}_{(i,j)} &= \sum_{l_a} \sum_{l_b} \mathbf{S}_{(i, l_a)} {\mathbf{G}}_{(l_a, l_b)} \mathbf{T}_{(l_b, j)} \\
    {\mathbf{M}_\alpha}_{(i,j)} &= \sum_{l_a} \sum_{l_b} \mathbf{S}_{(i, l_a)} {\mathbf{G}_\alpha}_{(l_a, l_b)} \mathbf{T}_{(l_b, j)}.
\end{align*}
Therefore, $\mathbb{E}|\mathbf{M}_{(i, j)}|^2$ and $\mathbb{E}|{\mathbf{M}_\alpha}_{(i, j)}|^2$ are calculated as follows:
\begin{align}
    &\mathbb{E}|\mathbf{M}_{(i,j)}|^2 \nonumber \\
    &= \sum_{l_a,l_{a'},l_b,l_{b'}} |\mathbf{S}_{(i,l_a)}\mathbf{S}_{(i,l_{a'})}\mathbf{T}_{(l_b,j)}\mathbf{T}_{(l_{b'},j)}| \times \mathbb{E}\left({\mathbf{G}}_{(l_a,l_b)}{\mathbf{G}}_{(l_{a'},l_{b'})}\right) \\
    \label{eq:em-1}
    &=\sum_{l_a} \sum_{l_b} |\mathbf{S}_{(i, l_a)}|^2 \cdot |\mathbf{T}_{(l_b, j)}|^2, \\
    &\mathbb{E}|{\mathbf{M}_\alpha}_{(i,j)}|^2 \nonumber\\
    &= \sum_{l_a,l_{a'},l_b,l_{b'}} |\mathbf{S}_{(i,l_a)}\mathbf{S}_{(i,l_{a'})}\mathbf{T}_{(l_b,j)}\mathbf{T}_{(l_{b'},j)}| \times \mathbb{E}\left({\mathbf{G}_\alpha}_{(l_a,l_b)}{\mathbf{G}_\alpha}_{(l_{a'},l_{b'})}\right) \\
    \label{eq:em-2}
    &= \alpha \sum_{l_a} \sum_{l_b} |\mathbf{S}_{(i, l_a)}|^2 \cdot |\mathbf{T}_{(l_b, j)}|^2 = \alpha \mathbb{E}|\mathbf{M}_{(i, j)}|^2.
\end{align}
We use the following facts for obtaining Eq. (\ref{eq:em-1}) and (\ref{eq:em-2}), respectively, that are led from the definition of variance and the product of two values from independent distributions.
\begin{align}
    \mathbb{E}\left(\mathbf{G}_{(i_1, j_1)}\mathbf{G}_{(i_2, j_2)}\right)&=\left\{
    \begin{array}{ll}
    1 & (i_1 = i_2\text{\&} j_1 = j_2) \\
    0 & \text{otherwise}
    \end{array}
    \right. \\
    \label{eq:Egg}
    \mathbb{E}\left({\mathbf{G}_\alpha}_{(i_1, j_1)}{\mathbf{G}_\alpha}_{(i_2, j_2)}\right)&= \left\{
    \begin{array}{ll}
    \alpha & (i_1 = i_2\text{\&} j_1 = j_2) \\
    0 & \text{otherwise}
    \end{array}
    \right.
\end{align}
\begin{flushright}
$\square$
\end{flushright}

\textit{Proof of Theorem \ref{thm:g-inv-fp}.}
For a Gaussian random matrix $\mathbf{G} \in \mathbb{R}^{m \times n}$, $(\mathbf{GG}^\top)^{-1}$ follows an inverse Wishart distribution \cite{muirhead_aspects_1982}.
The mean of the inverse Wishart distribution is calculated as follows:
\begin{equation}
    \mathbb{E}((\mathbf{GG}^\top)^{-1}) = 1/(n-m-1) \cdot \mathbf{S}_\mathbf{G}^{-1},
\end{equation}
where $\mathbf{S}_\mathbf{G}$ is the covariance matrix of $\mathbf{G}$.
Since $||\mathbf{G}^{\dag}||_F^2 = \text{tr}[(\mathbf{G}^\dag)^\top \mathbf{G}^\dag] = \text{tr}[(\mathbf{GG}^\top)^{-1}]$,
\begin{equation}
    \label{eq:eg}
    \mathbb{E}||\mathbb{\mathbf{G}^\dag}||_F^2 = 1/(n-m-1) \cdot \text{tr}[\mathbf{S}_{\mathbf{G}}^{-1}].
\end{equation}
When $\mathbf{G}$ is a $k \times (k-p)$ matrix $\mathbf{G}_r^\text{Y}$, the off-diagonal elements of $\mathbf{S}_\mathbf{GG}^{\top}$ are zeros and diagonal elements are $\alpha$.
Therefore, we obtain the following equation:
\begin{equation}
    \mathbb{E}||\mathbb{\mathbf{G}^\dag}||_F^2 = k/(p-1) \cdot \alpha^{-1}.
\end{equation}

\begin{flushright}
$\square$
\end{flushright}

\subsection{Numerical experiment}
\label{sec:ggta-eval}
\begin{figure}
    \centering
    \includegraphics[width=0.9\linewidth]{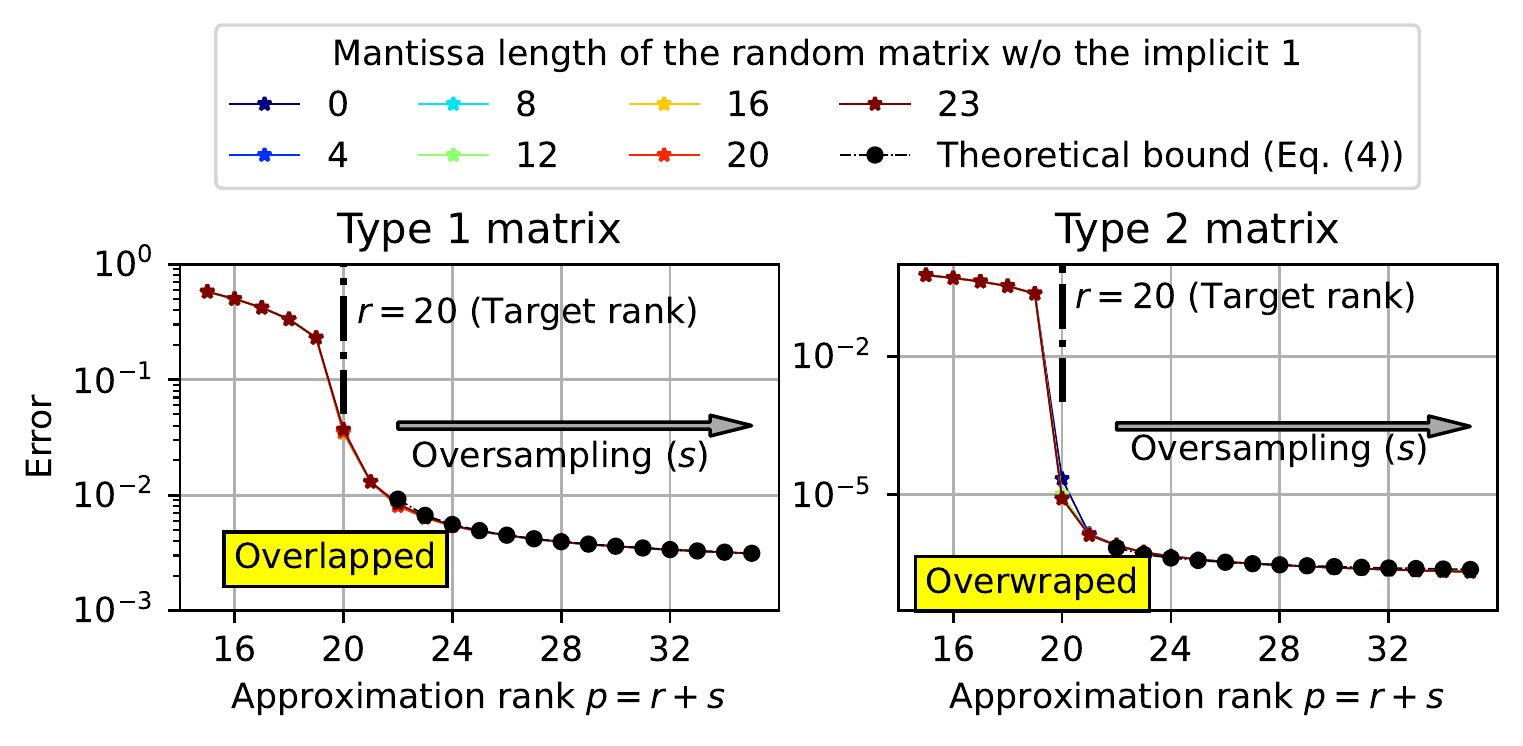}
    \caption{
    The accuracy of the random projection using low-precision Gaussian matrices for two kinds of input matrices.
    The error is calculated as $||\mathbf{A}-\mathbf{Q}\mathbf{Q}^\top\mathbf{A}||_F$.
    }
    \label{fig:mprp-accuracy}
\end{figure}
We conduct a numerical experiment to investigate the effect of the mantissa length on the accuracy of the random projection.
The FP32 input matrices $\mathbf{A}_{t1}$ and $\mathbf{A}_{t2}$ are generated in two ways that are used in the numerical experiment of \cite{connolly_randomized_2022} as follows.
\begin{itemize}
    \item Type 1:
    For $\mathbf{D} = \text{diag}(\mathbf{I}_r, 0) \in \mathbb{R}^{n \times n}$, we define
    \[
    \mathbf{A}_{t1} = \mathbf{D} + (\xi/\epsilon)\mathbf{GG}^\top,
    \]
    where $\mathbf{G}$ is a Gaussian random matrix. We fix $\xi = 10^{-4}, n=4096 $ and $ r=20$.
    \item Type 2: For orthogonal random matrices $\mathbf{U}, \mathbf{V} \in \mathbf{R}^{n \times n}$ in the Haar distribution, and
    \[
    \mathbf{D}=\text{diag}(\phi \mathbf{I}, 2^{-\alpha}, 3^{-\alpha}, \cdots, (n-r+1)^{-\alpha}),
    \]
    we define
    $\mathbf{A}_{t2} = \mathbf{U}\mathbf{D}\mathbf{V}^\top$.
    We fix $n=4096, r=20, \alpha=3$ and $\phi=10^6$.
\end{itemize}
The Gaussian random matrix $\mathbf{G}$ is generated in FP32 and rounded to low mantissa length values by RN.
To avoid the effect of the rounding error, we convert both the input matrices and random matrices to FP64 and compute the random projection in FP64.
The evaluation result is shown in Fig. \ref{fig:mprp-accuracy}.
In both test matrices, the accuracy of the random projection is almost the same across all mantissa lengths.

\subsection{Combination with existing sparse random projection methods}
The existing sparse random projection method uses a sparse random matrix generated by Eq. (\ref{eq:sp-rand}).
When computing the random projection using the sparse random matrix $\mathbf{\Omega}$, we do not need to multiply $\sqrt{n}$ in Eq. (\ref{eq:sp-rand}) since we only use the orthonormal matrix of the projected matrix.
Therefore, we can use a sparse random matrix where each element is one of $\{-1, 0, 1\}$ for the random projection.
These values can be represented even if the mantissa has only the implicit 1 bit.
Thus, the mixed-precision random projection method can also be applied to (very) sparse random matrix methods with the same accuracy.
However, we only focus on the Gaussian random projection in the rest of this paper since it is the best-understood class of the random matrices in RandNLA theoretically \cite{halko_finding_2011}.

\section{Single and half-precision matrix multiplication on Tensor Cores}
\subsection{Algorithm}
We can compute a multiplication of single-precision matrix $\mathbf{A}_\text{F32}$ and half-precision matrix $\mathbf{B}_\text{F16}$ in single-precision accuracy by just omitting a term in TCEC-SGEMM emulation method as follows:
\begin{align}
    \label{eq:shgemm-1}
    \mathbf{A}_\text{low} &\leftarrow \text{toLow}\left(\mathbf{A}_\text{F32}\right) \\
    \label{eq:shgemm-2}
    \Delta \mathbf{A}_\text{low} &\leftarrow \text{toLow}\left(\left(\mathbf{A}_\text{F32} - \text{toF32}\left(\mathbf{A}_\text{low}\right)\right)\times 2^{11}\right) \\
    \mathbf{B}_\text{low} &\leftarrow \text{toLow}\left(\mathbf{B}_\text{F16}\right) \\
    \label{eq:shgemm-3}
    \mathbf{C}_\text{SH} &\sim \mathbf{A}_\text{low}\cdot\mathbf{B}_\text{low} + \Delta\mathbf{A}_\text{low}\cdot\mathbf{B}_\text{low}\times2^{-11}.
\end{align}
Since the computational complexity of this method is $2/3$ of TCEC-SGEMM, it can compute the matrix multiplication faster when the matrix $\mathbf{B}_\text{F32}$ can be expressed by FP16.
We call this method {\bf SHGEMM} (Single- and Half-precision GEMM).

\subsection{Implementation}
\begin{figure}[t]
    \centering
    \includegraphics[width=0.7\linewidth]{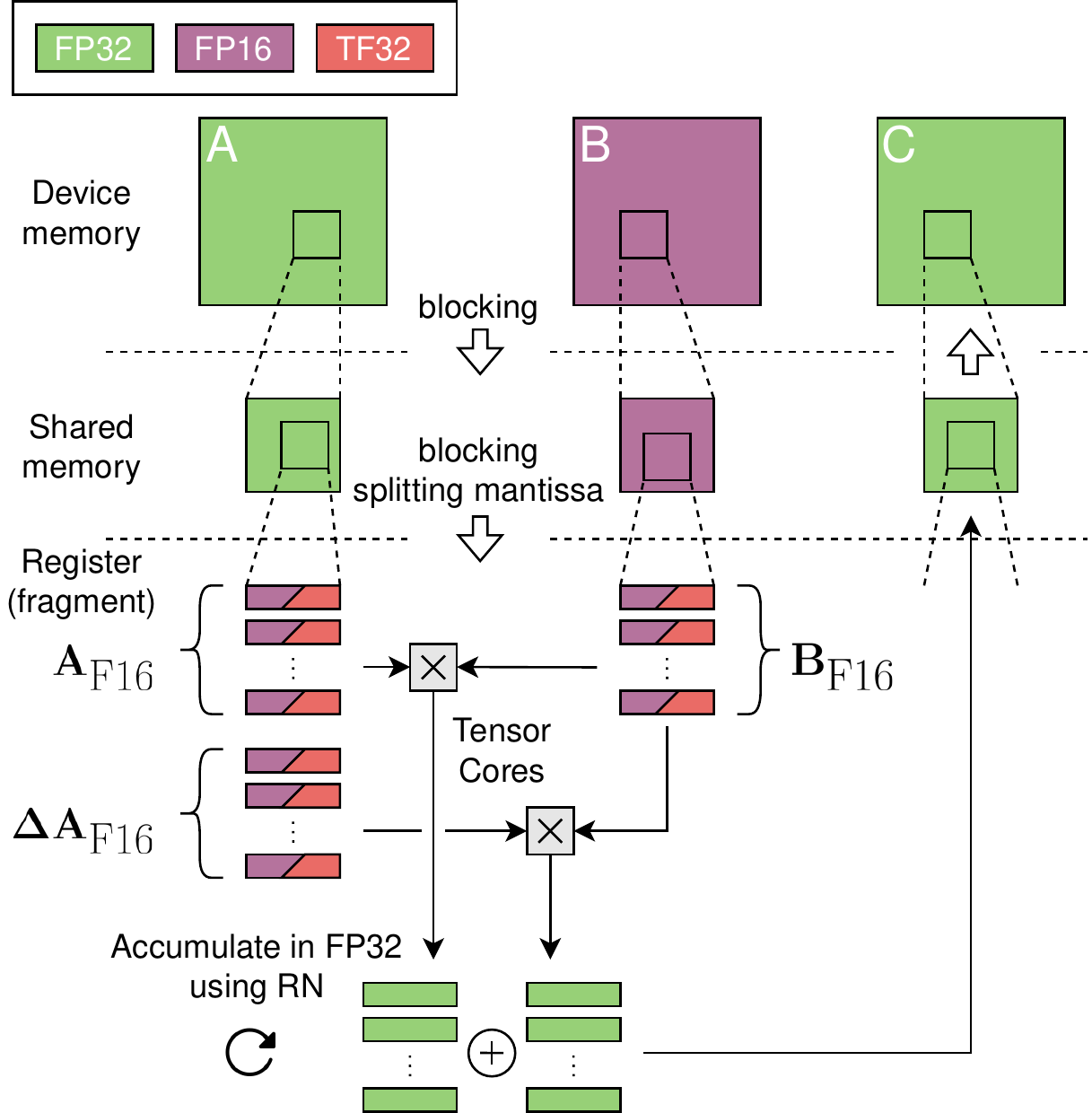}
    \caption{The computing flow of SHGEMM on Tensor Cores in our implementation.}
    \label{fig:shgemm-flow}
\end{figure}
We implement the SHGEMM kernel function from scratch using WMMA API and WMMA API extension library \cite{ootomo_reducing_2023}.
When using FP16 Tensor Cores, the supported exponent range of $\mathbf{A}_\text{F32}$ is limited to that of FP16.
Therefore, we implement two kinds of SGEMM: 1) limited exponent range SHGEMM using FP16 Tensor Cores (SHGEMM-FP16), and 2) full exponent range supporting SHGEMM using TF32 Tensor Cores (SHGEMM-TF32).
Since the theoretical peak performance of FP16 Tensor Cores is $2\times$ higher than TF32 Tensor Cores, SHGEMM-FP16 is theoretically faster than SHGEMM-TF32.
Even for the implementation of SHGEMM-TF32, the matrix $\mathbf{B}_\text{F16}$ is stored in FP16 on device memory, loaded and stored in FP16 on shared memory, and converted to TF32 on registers before input to Tensor Cores, as shown in Fig. \ref{fig:shgemm-flow}.
The implementation is available on GitHub\footnote{\url{https://github.com/enp1s0/shgemm}}.

\subsection{Rounding error analysis}
We analyze the rounding error bound of SHGEMM briefly and show that it is the same as the computation on FP32 SIMT Cores.
We use the {\tt mma.m16n8k8} PTX instruction which computes a matrix multiplication and addition on Tensor Cores.
It suffices to analyze the error bound of an inner product $c=\mathbf{x}^\top \mathbf{y} = \Sigma_{i=1}^{n} x_i y_i$.
The properties of the computation using the PTX instruction and the avoidance of RZ are as follows:
\begin{itemize}
    \item One instruction accumulates eight resulting values of the multiplication.
    \item The mantissa length of the accumulator inside Tensor Cores is 25-bit, including the implicit bit.
    \item A multiplication of two 11-bit mantissa values is conducted without rounding error and accumulated with RZ.
    \item The accumulated eight values are rounded by RZ and output in FP32.
    \item The accumulation of the resulting values of eight values is conducted on FP32 SIMT Cores with RN.
\end{itemize}
These properties have been confirmed theoretically or empirically through small numerical experiments.
The inner product is split into the accumulation of the eight elements accumulation as follows:
\begin{equation*}
    \label{eq:cti}
    c = \sum_{i=1}^{n'} t_i \text{ \ \ \  for \ \ \  } t_i = \sum_{j=1+8(i-1)}^{8i} x_j y_j,
\end{equation*}
where $n'=n/8$.
First, we obtain the rounding error bound of an inner product of two FP16 vectors.
We define $\tilde{c}$ and $\tilde{t}_i$ as the results of computing $c$ and $t_i$ on Tensor Cores, respectively, as follows:
\begin{equation*}
    \tilde{c} = \text{fl}_{RX}\left(\sum_{i=1}^{n'} \text{RZ}_{F32}\left(\tilde{t}_i\right)\right), \text{ \ \ } \tilde{t}_i = \text{fl}_{tc}\left(\sum_{j=1+8(i-1)}^{8i} x_j y_j\right),
\end{equation*}
where $\text{fl}_{RX}(f)$ denotes the computations of $f$ with RX for the rounding to be either RN or RZ. $\text{fl}_{tc}(f)$ is on Tensor Cores, which have the properties above, and $\text{RZ}_{F32}(\cdot)$ is a rounding function from 25-bit mantissa to FP32 by RZ.
In $\text{fl}_{RX}(f)$, the bound of rounding error is as follows:
\begin{align}
    \label{eq:rounding-error}
    \text{RX}(a) = a(1+\delta), |\delta| \leq \hat{u} \equiv \left\{
\begin{array}{ll}
 u & \text{RX is RN} \\
 2u & \text{RX is RZ},
\end{array}
\right.
\end{align}
where $u$ is the unit round off of FP32 and $2^{-24}$.
The RZ mode is the default mode on Tensor Cores, and the RN mode is used in SHGEMM, as same as TCEC-SGEMM.
Furthermore, the rounding error of $\tilde{t}_i$ are bounded as follows:
\begin{align*}
    \tilde{t}_i = t_i + \Delta_i \text{ \ s.t. \ }
    |\Delta_i| \leq \underbrace{7}_\text{Num additions}\cdot \underbrace{2u'}_\text{RZ} \cdot \left(\sum_{j=1+8(i-1)}^{8i} |x_j y_j|\right),
\end{align*}
where $u'=2^{-25}$.
Then, the following rounding error equation is satisfied.
\[
\tilde{c} = \sum_{i=1}^{n'} \text{RZ}_{F32}(\tilde{t}_i) + \Delta = \sum_{i=1}^{n'} (t_i + \Delta_i)(1+\delta_i) + \Delta
\]
where $\delta_i$ is $\delta$ for $\tilde{t}_i$ in Eq. (\ref{eq:rounding-error}) and $\Delta$ is a value that satifies $|\Delta| \leq (n'-1)\hat{u} \sum_{i=1}^{n'} \text{RZ}_{F32}(\tilde{t}_i)$.
We obtain the error bound of the inner product as follows by expanding inequalities of $\delta_i$, $\Delta_i$, and $\Delta$.
\begin{align*}
    |\tilde{c}-c| = \left|\tilde{c}-\sum_{i=1}^{n'} t_i\right| \lesssim \left|\sum_{i=1}^{n'} (t_i\delta_i + \Delta_i (1+\delta_1))+\Delta\right| \lesssim \frac{1}{8}n\hat{u}|x^\top||y|,
\end{align*}
where $|x^\top||y|$ is the inner product of elementwise absolute value vectors of $x$ and $y$.
Then, we extend the analysis of the rounding error of the inner product to a multiplication of two FP16 matrices in FP32, $\mathbf{A}_\text{F16} \cdot \mathbf{B}_\text{F16}$, and obtain its error bound inequalities as follows:
\begin{align}
    \label{eq:tc-err-bound}
    |\tilde{\mathbf{C}}-\mathbf{A}_\text{F16}\cdot\mathbf{B}_\text{F16}| \lesssim
    \left\{
\begin{array}{ll}
\frac{1}{8}nu |\mathbf{A}_\text{F16}||\mathbf{B}_\text{F16}| & \text{RX is RN} \\
\frac{1}{4}nu |\mathbf{A}_\text{F16}||\mathbf{B}_\text{F16}| & \text{RX is RZ},
\end{array}
\right.
\end{align}
for $n \geq 8$, where $\tilde{\mathbf{C}}$ is the computation result and $|\cdot|$ is the element-wise absolute value.

Next, we obtain the error bound of SHGEMM using the error bound of Tensor Cores and avoidance of RZ above.
In Eq. (\ref{eq:shgemm-1}-\ref{eq:shgemm-2}), the matrix $\mathbf{A}_\text{F32}$ is split into three terms as follows:
\begin{equation}
    \mathbf{A}_\text{F32} = \mathbf{A}_\text{F16} + \Delta \mathbf{A}_\text{F16} + \mathbf{A}_\Delta,
\end{equation}
where $\mathbf{A}_\Delta$ is the loss of mantissa caused by the splitting, where its expected length is 0.25-bit \cite{ootomo_recovering_2022}. 
For the split matrices, the following inequalities are satisfied:
\begin{equation}
    \label{eq:split-matrix-ineq}
    |\mathbf{A}_\text{F16}| \leq (1+u_\text{F16}) |\mathbf{A}_\text{F32}|, |\Delta\mathbf{A}_\text{F16}| \leq u_\text{F16}|\mathbf{A}_\text{F32}|, |\mathbf{A}_\mathbf{\Delta}| \leq u_\text{F16}^2 |\mathbf{A}_\text{F32}|,
\end{equation}
where $u_\text{F16}$ is the unit round off of FP16 and $2^{-11}$.
We compute the SHGEMM as follows:
\begin{align}
    \tilde{\mathbf{C}}_\text{SH} &= \text{fl}\left(\mathbf{A}_\text{F16}\mathbf{B}_\text{F16} + \Delta\mathbf{A}_\text{F16}\mathbf{B}_\text{F16}\right) \\
    &= \mathbf{A}_\text{F16}\mathbf{B}_\text{F16} + \mathbf{\Delta}^{(1)} + \Delta\mathbf{A}_\text{F16}\mathbf{B}_\text{F16} + \mathbf{\Delta}^{(2)} \nonumber \\
    \label{eq:shgemm-ineq}
    &+ \delta (\mathbf{A}_\text{F16}\mathbf{B}_\text{F16} + \mathbf{\Delta}^{(1)} + \Delta\mathbf{A}_\text{F16}\mathbf{B}_\text{F16} + \mathbf{\Delta}^{(2)}),
\end{align}
where $\mathbf{\Delta}^{(1)}$ and $\mathbf{\Delta}^{(2)}$ are the rounding error in $\mathbf{A}_\text{F16}\mathbf{B}_\text{F16}$ and $\Delta\mathbf{A}_\text{F16}\mathbf{B}_\text{F16}$, respectively.
We use Oomoto's RZ avoiding method for computing $\mathbf{A}_\text{F16}\mathbf{B}_\text{F16}$, but not for $\Delta\mathbf{A}_\text{F16}\mathbf{B}_\text{F16}$.
Therefore, the following inequalities are satisfied by Eq. (\ref{eq:tc-err-bound}) and (\ref{eq:split-matrix-ineq}):
\begin{align}
    |\mathbf{\Delta}^{(1)}| &\lesssim \frac{1}{8}nu |\mathbf{A}_\text{F32}||\mathbf{B}_\text{F16}| \\
    |\mathbf{\Delta}^{(2)}| &\lesssim \frac{1}{4}nuu_\text{F16} |\mathbf{A}_\text{F32}||\mathbf{B}_\text{F16}|.
\end{align}
Thus, we obtain the following inequality on the rounding error bound of SHGEMM by expanding Eq. (\ref{eq:shgemm-ineq}):
\begin{align}
    |\tilde{\mathbf{C}}_\text{SH} - \mathbf{A}_\text{F32}\cdot\mathbf{B}_\text{F16}| \lesssim \frac{1}{8}nu |\mathbf{A}_\text{F32}||\mathbf{B}_\text{F16}|,
\end{align}
for $n\geq 8$.
When computing the multiplication $\mathbf{A}_\text{F32} \cdot\mathbf{B}_\text{F16}$ all on FP32 SIMT Cores with 8 elements blocking similarly, the primary bounding term is also $\frac{1}{8}nu |\mathbf{A}_\text{F32}||\mathbf{B}_\text{F16}|$.
Therefore, the error bound of SHGEMM is the same as the computation on FP32 SIMT Cores.

\subsection{Evaluation of SHGEMM}
\label{sec:shgemm-eval}
All numerical experiments are conducted on NVIDIA A100 SXM4 (40GB) GPU and AMD EPYC 7702 CPU.
\subsubsection{Accuracy}
\begin{figure}[t]
    \centering
    \includegraphics[width=0.9\linewidth]{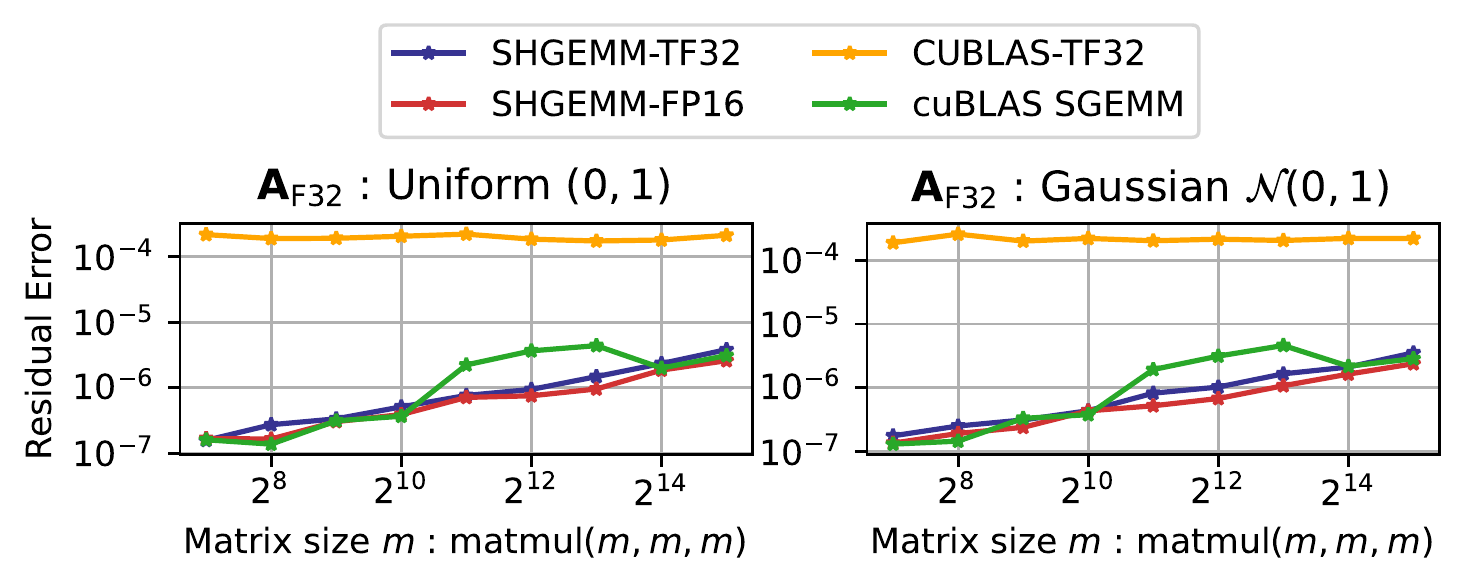}
    \caption{The accuracy of matrix multiplication $\mathbf{A}_\text{F32}\cdot\mathbf{B}_\text{F16}$ on SHGEMM. The matrix $\mathbf{B}_\text{F16}$ is generated with a Gaussian distribution $\mathcal{N}(0, 1)$ and $\mathbf{A}_\text{F32}$ is $\mathcal{N}(0, 1)$ (right) or a uniform distribution $(0, 1)$ (left).}
    \label{fig:shgemm-accuracy}
\end{figure}
The accuracy of the SHGEMM implementation is shown in Fig. \ref{fig:shgemm-accuracy}.
We compute the multiplication $\mathbf{A}_\text{F32} \cdot \mathbf{B}_\text{F16}$, where $\mathbf{A}_\text{F32}$ is initialized with $\mathcal{N}(0, 1)$ or a uniform distribution $(0, 1)$, and $\mathbf{B}_\text{F16}$ is $\mathcal{N}(0, 1)$.
We also compare the accuracy of cuBLAS SGEMM and cuBLAS TF32 GEMM by converting the matrix $\mathbf{B}_\text{F16}$ to FP32.
The relative error is calculated as follows:
\begin{equation*}
    \mathrm{Relative Error} = ||\mathbf{C}_\text{SH} - \mathbf{C}_\text{F64}||_F / ||\mathbf{C}_\text{F64}||_F,
\end{equation*}
where $\mathbf{C}_\text{F64}$ is computed in FP64 by converting the input matrices to FP64.
The accuracy of the SHGEMM implementations is at the same level as cuBLAS SGEMM.

\subsubsection{Throughput}
\begin{figure}[t]
    \centering
    \includegraphics[width=0.9\linewidth]{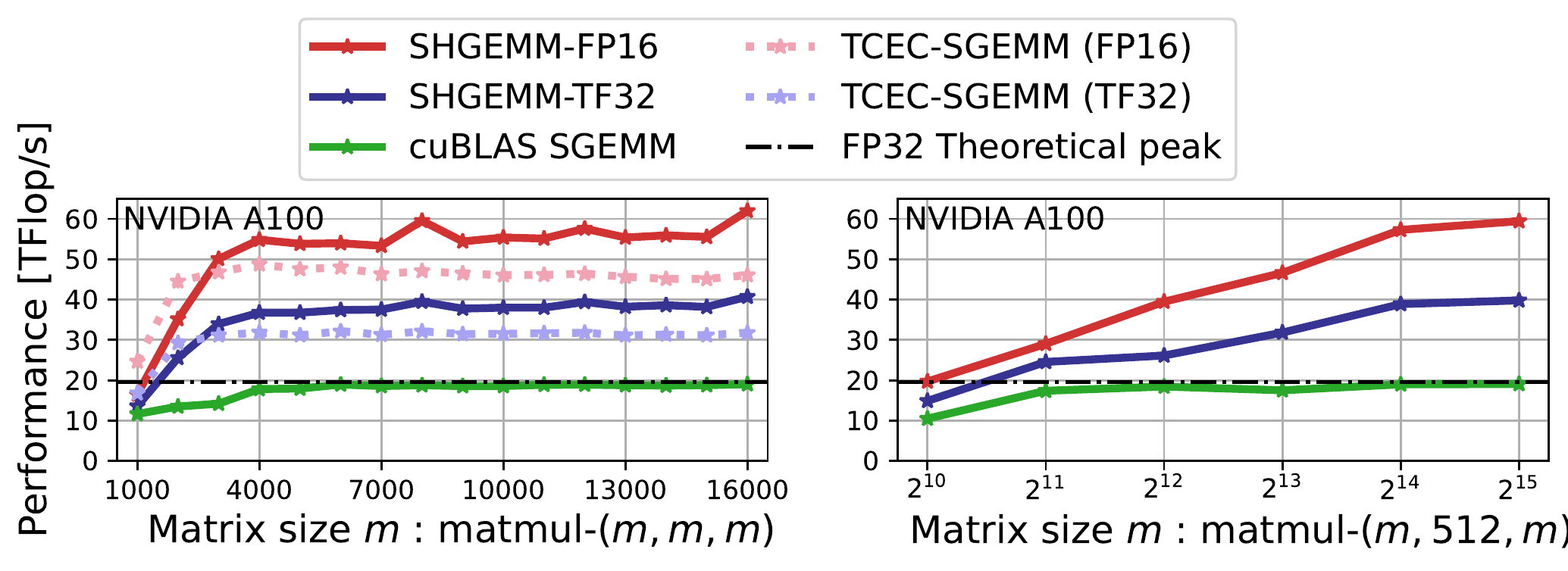}
    \caption{
    \textit{Left}: The throughput comparison of cuBLAS SGEMM, TCEC-SGEMM, and SHGEMM. The suffixes FP16 and TF32 denote the Tensor Core type used in the GEMM implementation.
    \textit{Right}: The throughput comparison of SHGEMM for tall-skinny matrix multiplication used in rank-$512$ Randomized SVD for $m \times m$ matrix.
    }
    \label{fig:shgemm-throughput}
\end{figure}
The throughput of the SHGEMM implementation is shown in Fig. \ref{fig:shgemm-throughput}.
The throughput is calculated as $2mnk/t$ [Flop/s] for matmul-($m, n, k$), which computes the multiplication of $m \times k$ and $k \times n$ matrices, and $t$ is the computing time [sec].
We compare the throughput of the two SHGEMM implementations, TCEC-SGEMM implementations that use FP16 Tensor Cores and TF32 Tensor Cores, and cuBLAS SGEMM.
The achieved throughput of SHGEMM-FP16 is up to $61.0$ [TFlop/s], and SHGEMM-TF32 is $40.5$ [TFlop/s].
Each SHGEMM implementation is faster than TCEC-SGEMM, which uses the same Tensor Cores.

The theoretical peak performance of SHGEMM-FP16 and SHGEMM-TF32 on NVIDIA A100 GPU are $312/2=156$ [TFlop/s] and $152/2=76$ [TFlop/s], respectively, under the following assumption.
\begin{assumption}
\label{ass:tc-every-clock}
All computations on computing units other than Tensor Cores can be overlapped with the ones on Tensor Cores.
In other words, the Tensor Core instruction is issued every clock cycle.
\end{assumption}
Therefore, the efficiency of SHGEMM-FP16 is only $39$\%, and SHGEMM-TF32 is $53$\%, respectively.
As the cause of the low efficiency, we consider it difficult to realize the Assumption \ref{ass:tc-every-clock} for the following reasons.

\begin{enumerate}
    \item
    There is a dependency in the order of Tensor Cores and FP32 SIMT Cores computations in Eq. (\ref{eq:shgemm-1}-\ref{eq:shgemm-3}).
    Therefore, these computations can not be overlapped.
    \item 
    The computing time on FP32 SIMT Cores for computing Eq. (\ref{eq:shgemm-2}) and avoiding RZ inside Tensor Cores is large.
    When splitting the resulting matrix of matmul($m, n, k$) into a grid where each matrix size is $b_m \times b_n$, and computing the result for each sub-divided matrices in parallel, the number of loaded $\mathbf{A}_\text{F32}$ elements is $mnk/b_n$.
    Therefore, the number of subtractions and multiplications in Eq. (\ref{eq:shgemm-2}) is $2mnk/b_n$, computed on FP32 SIMT Cores.
    Furthermore, we use FP32 SIMT Cores for accumulating the result of multiplications of sub-divided matrices to avoid RZ inside Tensor Cores $2mnk/f_k$ times, where $f_k$ is the $k$ value of the matmul that one Tensor Core instruction computes, which is $8$.
    On the other hand, the number of calculations on Tensor Cores in Eq. (\ref{eq:shgemm-3}) is $4mnk$.
    Thus, the computation ratio on FP32 SIMT Cores and Tensor Cores is $(2/b_n+1/f_k) : 4$ independent from $m, n$, and $k$.
    The ratio is about $1:25$ since we set $b_n=64$ or $128$ typically.
    Therefore, the computing time on FP32 SIMT Cores is not negligible against the one on Tensor Cores since the theoretical throughput ratio of FP32 SIMT Cores and Tensor Cores is only $1:16$ on FP16 Tensor Cores and $1:8$ on TF32 Tensor Cores.
\end{enumerate}

\section{Experiments}
\subsection{Randomized SVD}

\begin{figure*}[t]
    \centering
    \includegraphics[width=\linewidth]{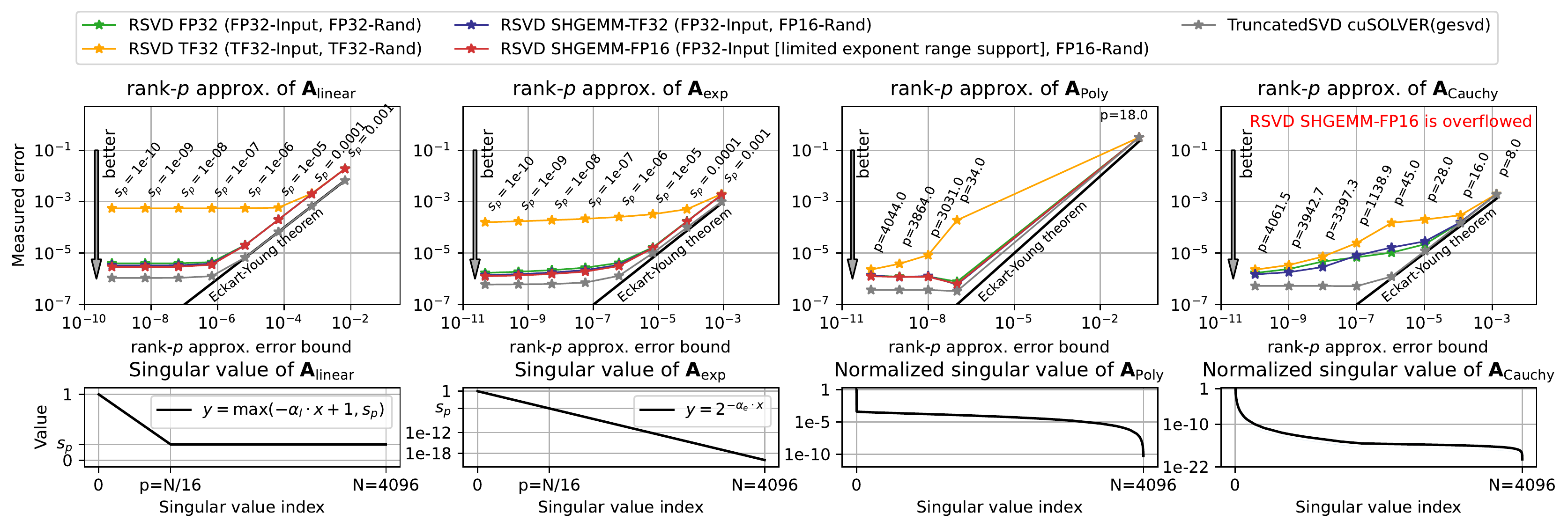}
    \caption{\textbf{Top:} The accuracy of Randomized SVD implementations for the input matrices $\mathbf{A}_\text{linear}, \mathbf{A}_\text{exp}$ and $\mathbf{A}_\text{Cauchy}$. \textbf{Bottom:} The singular value decay of input matrices and target ranks.}
    \label{fig:rsvd-eval}
\end{figure*}

We use SHGEMM in line 1 of Algorithm \ref{alg:rsvd} and NVIDIA cuBLAS and cuSOLVER libraries for the other part of the computations of the algorithm.

\subsubsection{Accuracy}
For the evaluation of accuracy, we use 4 kinds of test matrices as follows:
\begin{itemize}
    \item $\mathbf{A}_\text{linear}$ and $\mathbf{A}_\mathrm{exp} \in \text{FP32}^{4096 \times 4096}$:
    First, we decide a value $s_p \in (0, 1)$ and generate a FP32 value sequence $\{s_j\}$ as follows.
    \begin{itemize}
        \item For $\mathbf{A}_\text{linear}$: $s_i = \text{max}\left(-\alpha_l\times i + 1, s_p\right)$ where $\alpha_l=(1-s_p)/p$.
        \item For $\mathbf{A}_\text{exp}$: $s_i = 2^{-\alpha_e \times i}$ where $\alpha_e = (\log_2 1/s_p)/p$.
    \end{itemize}
    Then, we use the LAPACK {\tt slatms} function to make a $N \times N$ FP32 random matrix with $s_i$ as singular values.
    By generating the matrix in this way, we can calculate the error bound of $p$-rank approximation using Theorem \ref{the:eym}.
    Therefore, the error bound of relative residual of $p$-rank approximation for $\mathbf{A}_\text{linear}$ is
    \[
    s_p\sqrt{N-p}/||\mathbf{A}_\text{linear}||_F
    \]
    and for $\mathbf{A}_\text{exp}$ is
    \[
    \sqrt{(s_p^2-2^{2qN})/(1-2^{2q})}/||\mathbf{A}_\text{exp}||_F.
    \]
    In this evaluation, $N$ and $p$ are fixed to $4096$ and $256$.
    By fixing these parameters, we fix the computational complexity of Randomized SVD.
    Thus, the rounding error accumulation is also fixed to some extent, and we can compare the effect of the accuracy of random projection for the approximation accuracy.
    
    \item $\mathbf{A}_\text{Poly} \in \text{FP32}^{4096 \times 4096}$:
    The generation equation is the same as the matrix type 2 in \ref{sec:ggta-eval}.
    We used the parameters $r=20, \alpha=3$ and $\phi=10^6$.
    \item $\mathbf{A}_\text{Cauchy} \in \text{FP32}^{4096 \times 4096}$:
    We generate Cauchy matrices as follows and apply orthogonal iteration 1 time.
    \begin{equation*}
        {\mathbf{A}_\text{Cauchy}}_{i, j} = 1 / \left(|x_i - y_i| + \gamma\right),
    \end{equation*}
    where $x_i, y_i$ are random sequences in $(-10^{-3}, 10^{-3})$ and $\gamma = 10^{-3}$.
    The absolute value of some elements of this matrix exceeds the representation range of the FP16.
    Therefore, the computation is expected to fail when using SHGEMM-FP16.
\end{itemize}


We fix the oversampling parameter $s=10$ and evaluate ten matrices generated using different random seeds.
We compare the accuracy of the Randomized SVD across the GEMM implementations as shown in Fig. \ref{fig:shgemm-accuracy}.
The data type of the random matrix is FP16 when using SHGEMM, otherwise FP32.
The accuracy of RSVD TF32 degrades compared to RSVD FP32, which is the baseline, since the accuracy of TF32 GEMM is worse than SGEMM.
This implies that it is critical to preserve the accuracy of the input matrix during the random projection.
On the other hand, the accuracy of RSVD SHGEMM-TF32 and SHGEMM-FP16 are at the same level as RSVD FP32.

\subsubsection{Throughput}
\begin{figure*}[t]
    \centering
    \includegraphics[width=\linewidth]{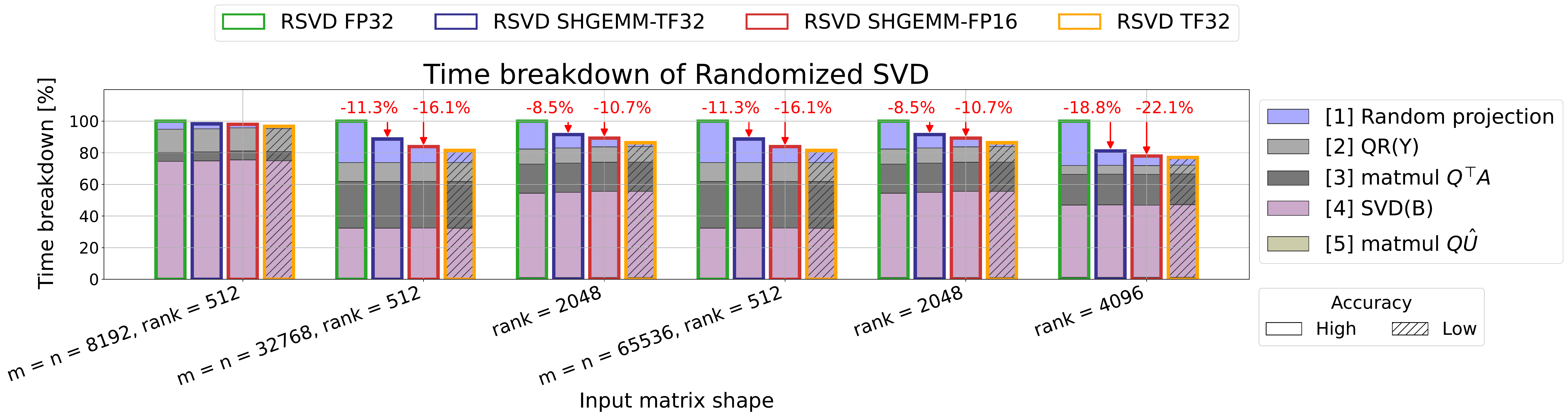}
    \caption{The time breakdown of baseline Randomized SVD (FP32) on NVIDIA A100 and computing time reduction by our method. The prefix numbers of the breakdown labels are the line numbers in Algorithm \ref{alg:rsvd}.}
    \label{fig:rsvd-time-breakdown}
\end{figure*}

The throughput comparison of the Randomized SVD implementations is shown in Fig. \ref{fig:rsvd-time-breakdown}.
Although the asymptotic computational complexity of the random projection is larger than SVD and QR decomposition, their real computing times are not negligible.
Therefore, in the case of $m=n=8192, rank=512$, we can not speed up the whole computation since their computing time consumes about 90\% of the whole.
On the other hand, we have reduced up to 18.9 \% and 22.1\% of the whole computing time using SHGEMM-TF32 and SHGEMM-FP16, respectively, while the accuracy is at the same level as FP32 when the computing time ratios of SVD and QR decomposition are not too large.

\subsection{Random Projection HOSVD}
We evaluate the effect of our method for High-Order Singular Value Decomposition (HOSVD) \cite{de_lathauwer_multilinear_2000}.
The HOSVD factorizes an $N$-dimension tensor $\mathbf{A} \in \mathbb{R}^{I_1 \times I_2 \times \cdots \times I_N}$ into multiplications of an $N$-dimensional tensor and matrices as follows:
\begin{equation}
    \mathbf{A} = \mathbf{g} \times_1 \mathbf{Q}_1^\top \times_2 \mathbf{Q}_2^\top \cdots \times_N \mathbf{Q}_N^\top,
\end{equation}
where $\mathbf{g}\in \mathbb{R}^{J_1 \times J_2 \times \cdots \times J_N}$ is a core tensor and $\mathbf{Q}_k \in \mathbb{R}^{I_k \times J_k}$ is an orthogonal matrix.
We denote the contraction of a tensor $\mathbf{T} \in \mathbb{R}^{I_1 \times I_2 \times \cdots \times I_N}$ and a matrix $\mathbf{M}\in\mathbb{R}^{I_i \times k}$ at $i$-th mode as $\mathbf{T} \times_i \mathbf{M}$.
The rank in each dimension determines the shape of the core tensor.
HOSVD is computed by flattening to matrix and SVD.
The random projection HOSVD (RP-HOSVD) \cite{ahmadi-asl_randomized_2021} shown in Algorithm \ref{alg:rhosvd} computes this factorization using random projection and QR factorization instead of SVD.
\begin{algorithm}[t]
\caption{Random projection HOSVD}\label{alg:rhosvd}
\begin{algorithmic}[1]
\Require Input matrix $\mathbf{A} \in \mathbb{R}^{I_1 \times I_2 \times \cdots \times I_N}$, Target rank $r \in\mathbb{N}$
\Ensure $\mathbf{g}\in \mathbb{R}^{J_1 \times J_2 \times \cdots \times J_N}, \mathbf{Q}_k \in \mathbb{R}^{I_k \times J_k}$ \text{ \ } s.t. $\mathbf{A} \sim \mathbf{g} \times_1 \mathbf{Q}_1 \times_2 \mathbf{Q}_2 \cdots \times_N \mathbf{Q}_N$
\For $ \text{ \ }i = 1..N$
\State $\mathbf{W} \leftarrow \mathbf{A}_{(i)}' \cdot \mathbf{\Omega}_{(i)}$ where $\mathbf{A}_{(i)}' \in \mathbb{R}^{I_i \times \left(\Pi_{k\neq i} I_k\right)}$ is a transposed matrix of the tensor $\mathbf{A}$, and $\mathbf{\Omega}_{(i)} \in \mathbb{R}^{\left(\Pi_{k\neq i} I_k\right) \times J_i}$ is a random matrix.
\State $\mathbf{Q}_i, \_ \leftarrow \text{QR}\left(\mathbf{W}\right)$
\EndFor
\State $\mathbf{g} \leftarrow \mathbf{A} \times_1 \mathbf{Q}^\top_1 \times_2 \mathbf{Q}^\top_2 \cdots \times_N \mathbf{Q}^\top_N$
\end{algorithmic}
\end{algorithm}

\begin{figure*}[t]
    \centering
    \includegraphics[width=\linewidth]{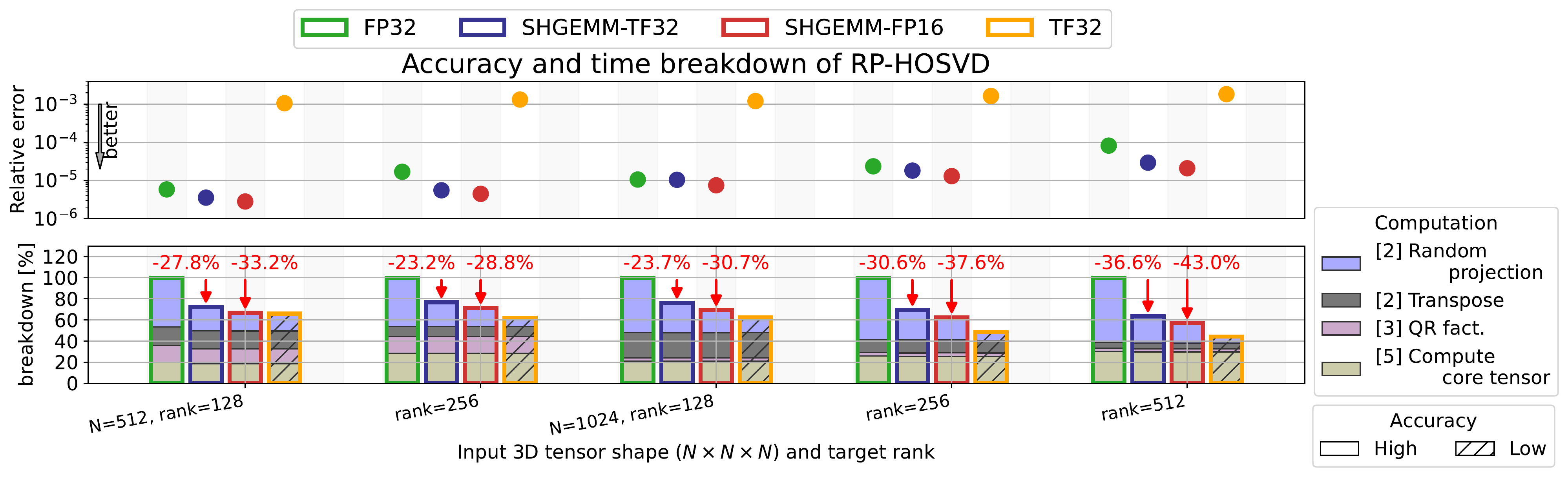}
    \caption{The accuracy (top) and time breakdown (bottom) of random projection HOSVD on NVIDIA A100. The prefix numbers of the breakdown labels are the line numbers in Algorithm \ref{alg:rhosvd}.}
    \label{fig:rphosvd-eval}
\end{figure*}

To evaluate RP-HOSVD, we generate test tensors as in Algorithm \ref{alg:rhosvd-input} and measure the approximation accuracy and throughput, as shown in Fig. \ref{fig:rphosvd-eval}.
As with the Randomized SVD accuracy, the accuracy degrades when we use TF32 random projection.
On the other hand, when using SHGEMM-TF32 and SHGEMM-FP16 random projection, we have reduced the computing times by 36.6\% and 43.0\%, respectively.
Although the accuracy of RP-HOSVD by SHGEMM is better than the FP32 baseline in some cases, we consider that the cause is the difference in the computation order, which is also shown in the unit test of SHGEMM in Fig. \ref{fig:shgemm-accuracy}.
Therefore, we conclude that the accuracy of RP-HOSVD by SHGEMM is almost at the same level as FP32.

\begin{algorithm}[t]
\caption{Tensor generation for RP-HOSVD evaluation}\label{alg:rhosvd-input}
\begin{algorithmic}[t]
\Require Input tensor dims ${I_i \in \mathbb{N}}$, Target ranks $J_i \in\mathbb{N}$, Rank padding $p \in \mathbb{N}$
\Ensure Input tensor $\mathbf{A} \in \mathbb{R}^{I_1 \times I_2 \times \cdots \times I_N}$
\State $\mathbf{G} \leftarrow$ GenRand($-1, 1$) $ \in \mathbb{R}^{J_1 \times J_2 \times \cdots \times J_N}$
\For $i = 1..N$
\State $\mathbf{\Omega}_\alpha \leftarrow$ GenRand($-1, 1$) $\in \mathbb{R}^{J_i \times (J_i - p)}$
\State $\mathbf{\Omega}_\beta \leftarrow$ GenRand($-1, 1$) $\in \mathbb{R}^{(J_i - p) \times I_i}$
\State $\mathbf{\Omega}_{(i)} \leftarrow \mathbf{\Omega}_\alpha \cdot \mathbf{\Omega}_\beta$ // This is a ($J_i-p$)-rank matrix
\State $\mathbf{G} \leftarrow \mathbf{G} \times_i \mathbf{\Omega}_{(i)}$
\EndFor
\end{algorithmic}
\end{algorithm}

\section{Conclusion}
We propose a fast mixed-precision random projection method for single-precision tensors leveraging the high throughput of Tensor Cores while maintaining approximation accuracy.
We confirm the property of low-precision Gaussian random values for the random projection.
The matrix multiplication for the random projection, SHGEMM, achieves $61.0$ TFlop/s using FP16 Tensor Cores and $40.5$ TFlop/s using TF32 Tensor Cores.
Using this method, we improve the throughput of Randomized SVD by $1.28$ times and Random projection HOSVD by $1.75$ times compared to baseline FP32 implementations while maintaining accuracy.

\subsection*{Acknowledgements}
The authors would like to thank Professor Katsuhisa Ozaki for technical assistance and discussion.
This work was supported by JSPS KAKENHI Grant Numbers JP20K20624, JP22H03598, and JP21J14694.
This work is supported by ”Joint Usage/Research Center for Interdisciplinary Large-scale Information Infrastructures” in Japan (Project ID: jh220022-NAH, jh220009-NAHI).

\bibliographystyle{plain} 
\bibliography{ref-short}

\end{document}